\pdfoutput=1

\documentclass[11pt,twoside,a4paper,cmspaper,final,collab]{cms-tdr}
\def\svnVersion{60779}\def\cmsCernNo{2011-002}\def\cmsCernDate{\today}\def\cmsMessage{Submitted to the Journal of High Energy Physics}

\begin{document}\cmsNoteHeader{EXO-10-013}

\hyphenation{had-ron-i-za-tion}
\hyphenation{cal-or-i-me-ter}
\hyphenation{de-vices}
\RCS$Revision: 43852 $
\RCS$HeadURL: svn+ssh://alverson@svn.cern.ch/reps/tdr2/papers/EXO-10-013/trunk/EXO-10-013.tex $
\RCS$Id: EXO-10-013.tex 43852 2011-03-04 20:56:26Z alverson $
\cmsNoteHeader{EXO-10-013} 

\newcommand{\numtrigs}{$10^{7}$}
\newcommand{\intlumi}{35\pbinv}
\newcommand{\energyresolutioneb}{1.0\%}
\newcommand{\energyresolutionee}{3.0\%}
\newcommand{\control}{$120 < M_{\ell\ell} < 200\GeV$}
\newcommand{\obseventscontrol}{XXX}
\newcommand{\expectedeventscontrolnonDY}{XXX}
\newcommand{\expectedeventscontrolDY}{XXX}
\newcommand{\fakescontrol}{YYY}
\newcommand{\efficelecEB}{90.1\% $\pm$ 0.5\%}
\newcommand{\efficelecEE}{87.2\% $\pm$ 0.9\%}
\newcommand{\efficSFEB}{0.979 $\pm$ 0.006}
\newcommand{\efficSFEE}{0.993 $\pm$ 0.011}
\newcommand{\SSMwidth}{30}
\newcommand{\Esixwidth}{6}
\newcommand{\Gonewidth}{14}
\newcommand{\Gtwowidt}{3.5}
\providecommand{\GKK}{\ensuremath{\mathrm{G}_\text{KK}}}
\providecommand{\tq}{\ensuremath{\mathrm{t}}}
\providecommand{\ZPSSM}{\ensuremath{\cPZpr_\text{SSM}}}
\providecommand{\ZPPSI}{\ensuremath{\cPZpr_\psi}}

\title{Search for Resonances in the Dilepton Mass Distribution in pp Collisions at $\sqrt{s}$ = 7 TeV}

\author[cern]{The CMS Collaboration}

\date{\today}

\abstract{

A search for narrow resonances at high mass in the dimuon and dielectron channels has
been performed by the CMS experiment at the CERN LHC, using  $\mathrm{pp}$
collision
data recorded at $\sqrt{s}=7$~TeV. The event samples correspond to
integrated luminosities of   $40~\mathrm{pb}^{-1}$ in the dimuon channel and
 $35~\mathrm{pb}^{-1} $ in the dielectron channel.
Heavy dilepton resonances are predicted in theoretical models with extra gauge
bosons ($\mathrm{Z}^\prime$)  or as Kaluza--Klein graviton excitations
($\mathrm{G}_\mathrm{KK}$) in the Randall-Sundrum model. Upper limits on the
inclusive cross section of
$\mathrm{Z}^\prime(\mathrm{G}_\mathrm{KK})\rightarrow\ell^+\ell^-$ relative to
$\mathrm{Z}\rightarrow\ell^+\ell^-$ are presented.
These limits exclude at 95\% confidence level
a $\mathrm{Z}^\prime$  with standard-model-like couplings
below 1140\GeV, the superstring-inspired $\mathrm{Z}^\prime_\psi$ below 887\GeV,
and, for values of the coupling parameter $k/\overline{M}_{\rm Pl}$ of
0.05 (0.1), Kaluza--Klein gravitons below 855 (1079)\GeV.
}

\hypersetup{%
pdfauthor={CMS Collaboration},%
pdftitle={Search for Resonances in the Dilepton Mass Distribution in pp Collisions at sqrt(s) = 7 TeV},%
pdfsubject={CMS},%
pdfkeywords={CMS, dileptons, resonances}}

\maketitle 

\newcommand{\fix}[1]{{\bf <<< #1 !!! }}%
\section{Introduction}

Many models of new physics predict the existence of narrow
resonances, possibly at the TeV mass scale, that decay to a pair of
charged leptons. This Letter describes a search for resonant signals
that can be detected by the Compact Muon Solenoid (CMS) detector at
the Large Hadron Collider (LHC)~\cite{lhc} at CERN. 
The Sequential Standard Model $\ZPSSM$ with standard-model-like couplings, the
$\ZPPSI$ predicted by grand unified theories~\cite{Leike:1998wr},
and Kaluza--Klein graviton excitations arising in 
the Randall-Sundrum (RS) model of extra
dimensions~\cite{Randall:1999vf, Randall:1999ee} were used as benchmarks. 
The RS model has two free parameters: the mass of the first graviton excitation and 
the coupling $k/\overline{M}_{\rm Pl}$, 
where $k$ is the curvature of the extra dimension and 
$\overline{M}_{\rm Pl}$ is the reduced effective Planck scale. 
Two values of the coupling parameter were considered: $k/\overline{M}_{\rm Pl}$~=~0.05 and 0.1.
For a resonance mass of 1\TeV, the widths are \SSMwidth, \Esixwidth\ and \Gtwowidt\ (\Gonewidth)\GeV 
for a $\ZPSSM$, $\ZPPSI$, and $\GKK$ with $k/\overline{M}_{\rm Pl}$~=~0.05 (0.1), respectively.

The results of searches for narrow $\cPZpr \rightarrow \ell^+ \ell^-$ and
$\GKK \rightarrow \ell^+ \ell^-$ resonances in $\Pp\Pap$ collisions at the
Tevatron with over
5\fbinv of integrated luminosity at centre-of-mass energy
of 1.96\TeV have previously
been reported~\cite{D0_RS,D0_Zp,CDF_RS,CDF_Zp}. 
Indirect constraints have been placed on the mass of the virtual 
$\cPZpr$ bosons by LEP-II experiments~\cite{delphi,aleph,opal,l3}
by examining the cross sections and angular distribution
of dileptons and hadronic final states in $\Pep\Pem$ collisions.

The results presented in this Letter were obtained from an analysis of
data recorded in 2010, corresponding to an integrated luminosity of $40 \pm
4$\pbinv in the dimuon channel, and $35 \pm 4$\pbinv in the dielectron
channel, obtained from $\Pp\Pp$ collisions at a centre-of-mass energy
of 7\TeV.  The total integrated luminosity used for the electron
analysis is smaller than that for the muon analysis because of the
tighter quality requirements imposed on the data.
The search for resonances is based on a shape analysis of dilepton mass spectra,
in order to be robust against uncertainties in the absolute background
level.  By examining the dilepton-mass spectrum from below the $\cPZ$
resonance to the highest mass events recorded, we obtain
limits on the ratio of the production cross section times branching
fraction for high-mass resonances to that of the $\cPZ$.  Using
further input describing the dilepton mass dependence on effects of
parton distribution functions (PDFs) and $k$-factors,
mass bounds are calculated for specific models. 
In addition, model-independent limit contours are determined in the
two-parameter $(c_d,c_u)$ plane~\cite{Carena:2004xs}. Selected
benchmark models for $\cPZpr$ production are illustrated in this plane,
where where $c_u$ and $c_d$ are model-dependent couplings of the
$\cPZpr$ to up- and down-type quarks, respectively allowing lower bounds 
to be determined.

\section{The CMS Detector}

The central feature of the CMS~\cite{JINST} apparatus is a
superconducting solenoid, of 6~m internal diameter, providing an axial field
of 3.8~T. Within the field volume are the silicon pixel and strip
trackers, the crystal electromagnetic calorimeter (ECAL) and the
brass/scintillator hadron calorimeter (HCAL). The endcap hadronic
calorimeters are segmented in the z-direction.  Muons are measured in
gas-ionization detectors embedded in the steel return yoke.  In
addition to the barrel and endcap detectors, CMS has extensive forward
calorimetry.

CMS uses a right-handed coordinate system, with the origin at the
nominal interaction point, the $x$-axis pointing to the centre of the
LHC, the $y$-axis pointing up (perpendicular to the LHC plane), and
the $z$-axis along the anticlockwise-beam direction. The polar angle,
$\theta$, is measured from the positive $z$-axis and the azimuthal
angle, $\phi$, is measured in the $x$-$y$ plane.

Muons are measured in the pseudorapidity range $|\eta|< 2.4$, with
detection planes based on one of three technologies: drift tubes in the barrel
region, cathode strip chambers in the endcaps, and resistive plate chambers
in the barrel and part of the endcaps.
The inner tracker (silicon pixels and strips) detects charged particles within the pseudorapidity range
$|\eta| < 2.5$.

The electromagnetic calorimeter consists of nearly 76\,000 lead
tungstate crystals which provide coverage in pseudorapidity $|\eta| < 1.479$
in the barrel region (EB, with crystal size
$\Delta\eta=0.0174$ and $\Delta\phi = 0.0174$) and $1.479 < |\eta| <
3.0$ in the two endcap regions (EE, with somewhat larger crystals.).
A preshower detector
consisting of two planes of silicon sensors interleaved with a total of
3$\,X_0$ of lead is located in front of the EE.

The first level (L1) of the CMS trigger system, composed of custom
hardware processors, selects the most interesting events
using information from the calorimeters and muon detectors.
The High Level Trigger (HLT) processor farm further decreases the
event rate employing the full event information, including the
inner tracker.  The muon selection algorithms in the HLT use information from the
muon detectors and the silicon pixel and strip trackers.  The electromagnetic (EM)
selection algorithms use the energy deposits in the ECAL and HCAL; the
electron selection in addition requires tracks matched to clusters. Events
with muons or electromagnetic clusters with $\pt$ above L1 and HLT
thresholds are recorded.

\section{Electron and Muon Selection \label{sec:lepton}}

\subsection{Triggers \label{sec:triggers}}

The events used in the dimuon channel analysis were collected using a
single-muon trigger. 
The algorithm requires a muon candidate to be found in the muon detectors by the L1 trigger.
The candidate track is then matched to a silicon tracker track, forming an HLT muon.
The HLT muon is required to have $\pt >9$ to 15\GeV, depending on the running period.  

A double EM cluster trigger was used to select
the events for the dielectron channel.  ECAL clusters are formed by
summing energy deposits in crystals surrounding a ``seed'' that is
locally the highest-energy crystal. 
The clustering algorithm takes into account the emission of bremsstrahlung.
This trigger requires two clusters with the ECAL transverse energy $\ET$
above a threshold of 17 to 22\GeV, depending on the running period. 
For each of these clusters, the ratio $H/E$, where $E$ is the energy of the ECAL cluster and
 $H$ is the energy in the HCAL cells situated behind it, is required to be less than 15\%.
At least one of these clusters must
have been associated with an energy deposit identified by the L1 trigger.

\subsection{Lepton Reconstruction\label{sec:leptonID}}

The reconstruction, identification, and calibration of muons
and electrons
follow standard CMS methods~\cite{EWK-10-002-PAS}.
Combinations of test beam, cosmic ray muons, and 
data from proton collisions 
have been used to calibrate the relevant detector systems
for both muons and electrons.

Muons are reconstructed independently as tracks in both the muon
detectors and the silicon tracker~\cite{MUONPAS}. 
The two tracks can be matched and fitted simultaneously to
form a ``global muon''. Both muons in the event must be identified as
global muons, with at least 10 hits in the silicon tracker and with
$\pt > 20$\GeV.  All muon candidates that satisfy these criteria
are classified as ``loose'' muons.  At least one of the two muons in
each event must be further classified as a ``tight'' muon by passing the
following additional requirements: a transverse impact parameter with
respect to the collision point less than 0.2~cm; a $\chi^2$ per degree
of freedom less than 10 for the global track fit; at least one hit in
the pixel detector; hits from the muon tracking system in at least two
muon stations on the track; and correspondence with the single-muon trigger.

Electrons are reconstructed by associating a cluster in the ECAL with
a track in the tracker~\cite{EGMPAS}.  Track reconstruction, which 
is specific to
electrons to account for bremsstrahlung emission,  is seeded from the
clusters in the ECAL, first using the cluster position and energy to
search for compatible hits in the pixel detector, and then using these
hits as seeds to reconstruct a track in the silicon tracker. A minimum
of five hits is required on each track.  Electron candidates are required to be
within the barrel or endcap acceptance regions, with pseudorapidities
of $|\eta|<1.442$ and $1.560<|\eta|<2.5$, respectively. A candidate
electron is required to deposit most of its energy in the ECAL and
relatively little in the HCAL ($H/E<5\%$). The transverse shape of the
energy deposit is required to be consistent with that expected for an
electron, and the associated track must be well-matched in $\eta$ and
$\phi$. Electron candidates must have $\ET > 25$~GeV.

In order to suppress misidentified leptons from jets and non-prompt muons from
hadron decays, both lepton selections impose isolation requirements.
Candidate leptons are required to be isolated within a narrow cone of
radius $\Delta R = \sqrt{(\Delta\eta)^2 + (\Delta\phi)^2} = 0.3$,
centred on the lepton. Muon isolation requires that the sum of the
$\pt$ of all tracks within the cone, excluding the muon,
is less than 10\% of the $\pt$ of the muon.
For electrons, the sum of the $\pt$ of the tracks, excluding
the tracks within an inner cone of $\Delta R = 0.04$, is required to be less than 7$\GeV$
for candidates reconstructed within the barrel acceptance and 15$\GeV$
within the endcap acceptance. The calorimeter isolation
requirement for electron candidates within the barrel acceptance is that,
excluding the $\ET$ of the candidate, the sum of the $\ET$ resulting from
deposits in the ECAL and the HCAL within a cone of $\Delta R=0.3$ be less than
0.03$\ET$ + 2\GeV. For candidates within the endcap acceptance, the
segmentation of the HCAL in the $z$-direction is exploited. For candidates with $\ET$
below 50\GeV (above 50\GeV), the isolation energy is required to be
less than 2.5\GeV ($0.03(\ET-50) + 2.5\GeV)$, where $\ET$ is
determined using the ECAL and the first layer of the segmented HCAL. The
$\ET$ in the second layer of the HCAL is required to be less than
0.5\GeV.  These requirements ensure that the candidate electrons are
well-measured and have minimal contamination from jets.

The performance of the detector systems for the data sample presented in this paper is 
established using measurements of standard model (SM) $\PW$ and $\cPZ$ processes 
with leptonic final states~\cite{EWK-10-002-PAS} 
and using traversing cosmic ray muons~\cite{CMS_CFT_09_014}.

Muon momentum resolution varies from 1\% at momenta
of a few tens of \GeV to 10\% at momenta of several hundred \GeV,
as verified with measurements made with cosmic rays.
The alignment of the muon and inner tracking systems is important
for obtaining the best momentum resolution, and hence mass resolution,
particularly at the high masses relevant to the $\cPZpr$ search.
An additional contribution to the momentum
resolution arises from the presence of distortion modes in the tracker
geometry that are not completely constrained by the alignment procedures.
The dimuon mass resolution is estimated to have an rms of
5.8\% at 500\GeV and 9.6\% at 1\TeV.

The ECAL has an ultimate energy resolution of better than $0.5\%$ for
unconverted photons with transverse energies above $100\GeV$.
The ECAL energy resolution obtained thus far is
on average 1.0\% for the barrel and 4.0\% for the endcaps.
The mass resolution is estimated to be
1.3\% at 500\GeV and 1.1\% at 1\TeV.
Electrons from $\PW$ and $\cPZ$ bosons were used to calibrate ECAL
energy measurements.  
For both muons and electrons, the energy scale
is set using the $\cPZ$ mass peak, except for electrons in the barrel
section of the ECAL, where the energy scale is set using neutral pions,
and then checked using the $\cPZ$ mass peak.
The ECAL energy scale uncertainty is 1\% in the barrel and 3\% in the
endcaps.

\subsection{Efficiency Estimation \label{sec:eff}}

The efficiency for identifying and reconstructing lepton candidates is
measured with the tag-and-probe method~\cite{EWK-10-002-PAS}.
A tag lepton is established by applying tight cuts
to one lepton candidate; the other candidate is used as a probe. A
large sample of high-purity probes is obtained by requiring that the
tag-and-probe pair have an invariant mass consistent with the $\cPZ$
boson mass ($80 < m_{\ell\ell} < 100\GeV)$.  
The factors contributing to the overall efficiency are measured in the data. They are:
the trigger efficiency, the reconstruction efficiency in the silicon tracker,
the electron clustering efficiency, and the lepton reconstruction and
identification efficiency. All efficiencies and scale factors
quoted below are computed
using events in the $\cPZ$ mass region.

The trigger efficiencies are defined relative to the full offline
lepton requirements.  For the dimuon events, the efficiency of the
single muon trigger with respect to loose muons is measured to be
$89\% \pm 2\%$~\cite{EWK-10-002-PAS}.  The overall efficiency, defined
with respect to particles within the physical acceptance of the
detector, for loose (tight) muons is measured to be $94.1\%\pm1.0\%$
($81.2\%\pm1.0\%$).  Within the statistical precision allowed by the
current data sample, the dimuon efficiency is constant as a function of
$\pt$ above 20\GeV, as is the ratio of the efficiency in the data to that in the Monte Carlo (MC)
of 0.977 $\pm$ 0.004. For dielectron events, the double EM cluster trigger is
100\% efficient (99\% during the early running period).  The total
electron identification efficiency is \efficelecEB\ (barrel) and \efficelecEE\ (endcap). 
The ratio of the electron efficiency measured from the data to that
determined from MC simulation at the $\cPZ$ resonance is \efficSFEB\ (EB)
and \efficSFEE\ (EE).  To determine the efficiency applicable to
high-energy electrons in the data sample, this correction factor is
applied to the efficiency found using MC simulation.
The efficiency of electron identification increases as a
function of the electron transverse energy until it becomes flat beyond an
$\ET$ value of about 45\GeV. Between 30 and 45 GeV it increases by about 5\%.

\section{Event Samples and Selection}

Simulated event samples for the signal and associated backgrounds
were generated with the \PYTHIA~{\sc v6.422}~\cite{Sjostrand:2006za}
MC event generator, and with
\MADGRAPH~\cite{MADGRAPH} and
{\sc powheg v1.1}~\cite{Alioli:2008gx, Nason:2004rx, Frixione:2007vw}
interfaced with the \PYTHIA parton-shower generator 
using the {\sc CTEQ6L1}~\cite{Pumplin:2002vw} PDF set.
The response of the detector was simulated in detail using
\GEANTfour~\cite{GEANT4}. These samples were further processed
through the trigger emulation and event reconstruction chain of the CMS
experiment.

For both dimuon and dielectron final states, 
two isolated same flavour leptons that pass
the lepton identification criteria described in Section~\ref{sec:leptonID} are required.
The two charges are
required to have opposite sign in the case of dimuons (for which a charge
misassignment implies a large momentum measurement error), but not in the
case of dielectrons (for which charge assignment is decoupled from
the ECAL-based energy measurement). 
An opposite-charge requirement for
dielectrons would lead to a loss of signal efficiency of a few percent.

Of the two muons selected, one is required to satisfy the ``tight''
criteria. 
The electron sample requires at least one
electron candidate in the barrel because events with both electrons in the endcaps
will have a lower signal-to-background ratio.  
For both channels, each event is required to have a
reconstructed vertex with at least four associated tracks, located
less than 2~cm from the centre of the detector in the direction
transverse to the beam and less than 24\cm in the direction along the
beam. This requirement provides protection against cosmic rays.
Additional suppression of cosmic ray muons
is obtained by requiring the
three-dimensional opening angle between the two muons to be smaller than $\pi - 0.02$ radians.

\section {Backgrounds}

The most prominent SM process that contributes to the dimuon and
dielectron invariant mass spectra is Drell--Yan production
($\cPZ{/}\gamma^*$);
there are also  contributions from the $\ttbar$, $\tq\PW$, $\PW\PW$,  and
$Z\rightarrow\tau\tau$ channels. In addition, jets may
be misidentified as leptons and contribute to the dilepton invariant
mass spectrum through multi-jet and vector boson + jet final states.

\subsection{\texorpdfstring{\cPZ/$\gamma^*$}{Z/gamma*} Backgrounds}

The shape of the dilepton invariant mass spectrum is obtained from
Drell--Yan production using a MC simulation based on the
\PYTHIA event generator.   The simulated
spectrum at the invariant mass peak of the $\cPZ$ boson is normalized to
the data.  The dimuon analysis uses the data events in the $\cPZ$
mass interval of 60--120\GeV; the dielectron analysis
uses data events in the narrower interval of 80--100\GeV in order to
obtain a comparably small background contamination.

A contribution to the uncertainty attributed to the extrapolation of the
event yield and the shape of the Drell--Yan background to high
invariant masses arises from higher order QCD corrections.
The next-to-next-to-leading order (NNLO) $k$-factor is computed using
{\sc FEWZz v1.X}~\cite{Melnikov:2006kv}, with \PYTHIA~{\sc v6.409} and
{\sc CTEQ6.1} PDF~\cite{Stump:2003yu}
as a baseline. It is found that the variation of the
$k$-factor with mass does not exceed 4\% where
the main difference arises from the comparison of \PYTHIA and {\sc FEWZz} calculations.
A further source of uncertainty arises from the PDFs. The  {\sc lhaglue}~\cite{Bourilkov:2003kk}
interface to the {\sc lhapdf-5.3.1}~\cite{Whalley:2005nh} library is used to evaluate these uncertainties,
using the error PDFs from the {\sc CTEQ6.1} and the MRST2006nnlo~\cite{Martin:2007bv}
uncertainty eigenvector sets.
The uncertainty on the ratio of the background in
the high-mass region to that in the region of the $\cPZ$ peak is below
4\% for both PDF sets and masses below 1 TeV. Combining the higher order QCD and PDF uncertainties in
quadrature, the resulting uncertainty in the number of events normalized to those expected at the
$\cPZ$ peak is about 5.7\% for masses between 200\GeV and 1\TeV.

\subsection{Other Backgrounds with Prompt Lepton Pairs
\label{sec:e-mu}  }

The dominant non-Drell--Yan electroweak contribution to high
$m_{\ell\ell}$ masses is {$\ttbar$}; in addition there are contributions
from $\tq\PW$ and diboson production. In the $\cPZ$ peak
region, $\cPZ \rightarrow \Pgt\Pgt$ decays also contribute.
All these processes are flavour symmetric and produce twice as
many $\Pe\Pgm$ pairs as $\Pe\Pe$ or $\Pgm\Pgm$ pairs.
The invariant mass spectrum from
$\Pe^\pm\Pgm^\mp$ events is expected to have the same shape as that of same
flavour $\ell^+\ell^-$ events but without significant contamination
from Drell--Yan production.

\begin{figure}
\begin{center}
\includegraphics[angle=90,width=0.49\textwidth]{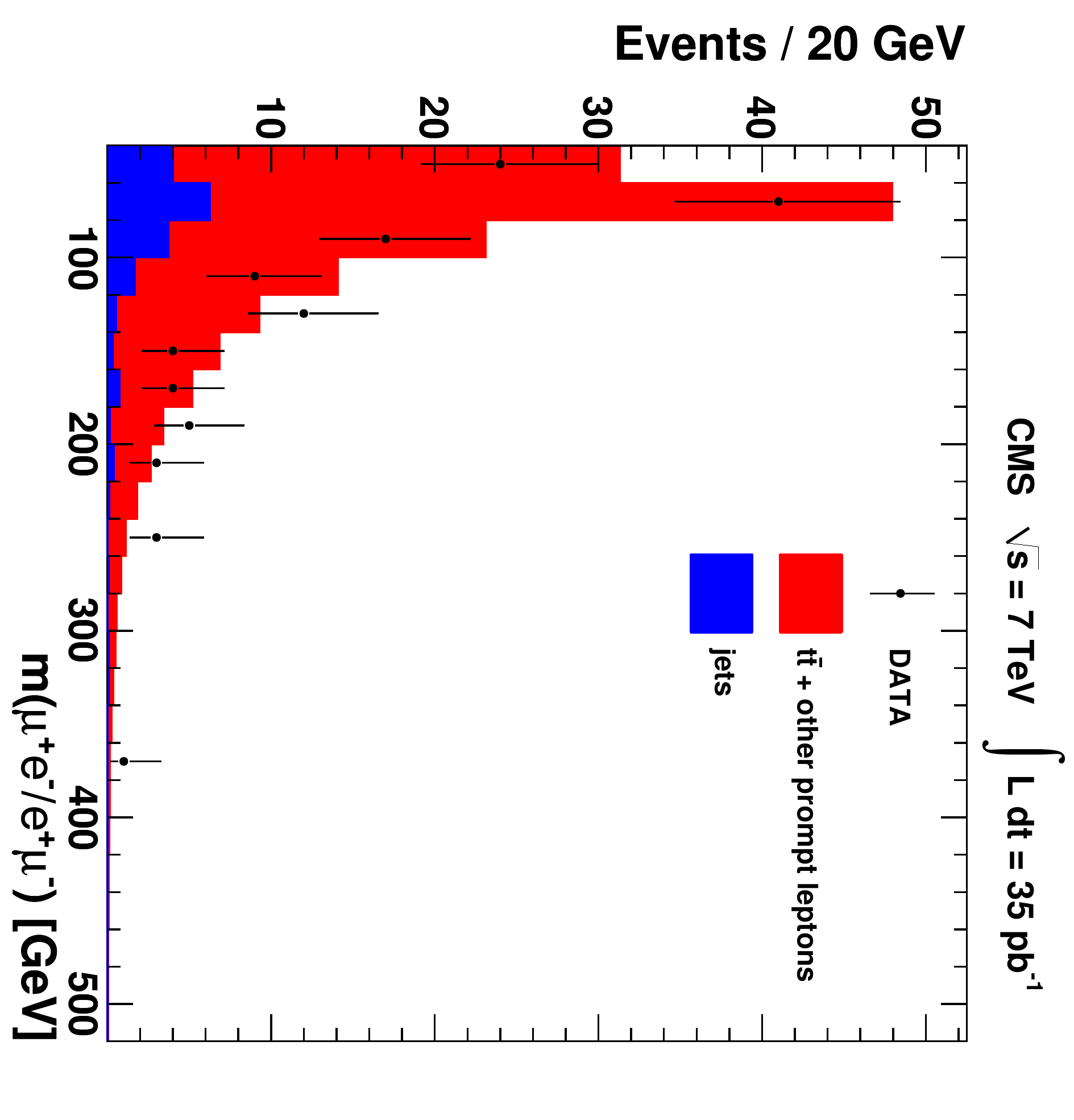}
\end{center}
\caption{\label{fig:muonselectrons}
The observed opposite-sign $\Pe^\pm\Pgm^\mp$ dilepton invariant mass spectrum
(data points). The uncertainties on the data points (statistical only)  
represent 68\% confidence intervals for the Poisson means.
Filled histograms show contributions to the
spectrum from  $\ttbar$, other sources of prompt leptons
 ($\tq\PW$, diboson production, $\cPZ\to\Pgt\Pgt$),  and
the multi-jet background (from Monte Carlo simulation).
}
\end{figure}

Figure~\ref{fig:muonselectrons} shows the
observed $\Pe^\pm\Pgm^\mp$ dilepton invariant mass spectrum
from a dataset corresponding to 35\pbinv, overlaid on
the prediction from simulated background processes. This spectrum was
obtained using the same single-muon trigger as in the dimuon analysis
and by requiring oppositely charged leptons of different flavour. Using an electron
trigger, a very similar spectrum is produced.
Differences in the geometric acceptances and efficiencies result in the predicted ratios
of $\Pgmp\Pgmm$ and $\Pe\Pe$ to $\Pe^\pm\Pgm^\mp$ being approximately 0.64
and 0.50, respectively. In the data, shown in
Fig.~\ref{fig:muonselectrons}, there are 32 (7) $\Pe^\pm\Pgm^\mp$ events
with invariant mass above 120 (200)\GeV. This yields an expectation of
about 20 (4) dimuon events and 16 (4) dielectron events.  A direct
estimate from MC simulations of the processes involved
predicts $20.1\pm3.6$ $(5.3\pm 1.0)$ dimuon events and
$13.2\pm2.4$ $(3.5\pm0.6)$ dielectron events. The uncertainty
includes  both statistical and systematic contributions, and is dominated
by the theoretical uncertainty  of 15\% on the $\ttbar$ production cross
section~\cite{Campbell:2010ff,Kleiss:1988xr}.
The agreement
between the observed and predicted distributions provides a validation
of the estimated contributions from the backgrounds from prompt leptons
obtained using MC simulations.

\subsection{Events with Misidentified and Non-Prompt Leptons}

A further source of background arises when objects are falsely
identified as prompt leptons.  The misidentification of jets as
leptons, the principle source of such backgrounds, is more likely to
occur for electrons than for muons.

Backgrounds arising from jets that are misidentified as electrons
include $\PW\to \Pe\cPgn$ + jet events with
one jet misidentified as a prompt electron, and also multi-jet events with
two jets misidentified as prompt electrons.
A prescaled single EM cluster trigger is used for collecting a sample of
events to determine the rate of jets misreconstructed as electrons and
to estimate the backgrounds from misidentified electrons.
The events in this sample are
required to have no more than one
reconstructed electron, and missing transverse energy of less than 20\GeV,
to suppress the contribution from $\cPZ$
and $\PW$ events respectively.
The probability for an EM cluster with $H/E<5\%$ to be
reconstructed as an electron is determined in bins of $\ET$ and $\eta$ from
a data sample dominated by multi-jet events and is used to weight
appropriately  events which have two such clusters passing
the double EM trigger.
The estimated background
contribution to the dielectron mass spectrum due to misidentified jets is
 8.6$\pm$3.4 (2.1$\pm$0.8)  for $m_{\Pe\Pe} > 120$ (200)\GeV.

In order to estimate the residual contribution from background events
with at least one non-prompt or misidentified muon, events are
selected from the data sample with single muons that pass all
selection cuts except the isolation requirement.
A map is created, showing the isolation probability for these muons as
a function of $\pt$ and $\eta$.
This probability map is corrected for the expected
contribution from events with single prompt muons from $\ttbar$ and
$\PW$ decays and for the observed correlation between the
probabilities for two muons in the same event. The probability map is
used to predict the number of background events with two isolated
muons based on the sample of events that have two non-isolated muons.
This procedure, which has been validated using simulated events,
predicts a mean background to $m_{\Pgm\Pgm} > 120$ (200)\GeV of
$0.8\pm0.2\ (0.2\pm0.1)$.

As the signal sample includes the requirement that the muons in the
pair have opposite electric charge, a further cross-check of the
estimate is performed using events with two isolated muons of the same
charge.  There are
no events with same-charge muon pairs and $m_{\Pgm\Pgm} > 120$\GeV,
a result which is statistically compatible with both the figure of
 $1.6\pm0.3$ events predicted from SM process using MC simulation and
the figure of $0.4\pm0.1$ events obtained using methods based on data.

\subsection{Cosmic Ray Muon Backgrounds}

The $\Pgmp\Pgmm$ data sample is susceptible to contamination from
traversing cosmic ray muons, which may  be misreconstructed as a pair
of oppositely charged, high-momentum muons.
Cosmic ray events can be removed from the data sample
because of their distinct topology (collinearity of two tracks
associated with  the
same muon),
and their uniform distribution of impact parameters with respect to the collision vertex.
The residual mean expected background from cosmic ray muons is
measured using sidebands to be less than 0.1 events with
$m_{\Pgm\Pgm} > 120$\GeV.

\section{Dilepton Invariant Mass Spectra}

\begin{figure}
\begin{center}
\includegraphics[angle=90,width=0.49\textwidth]{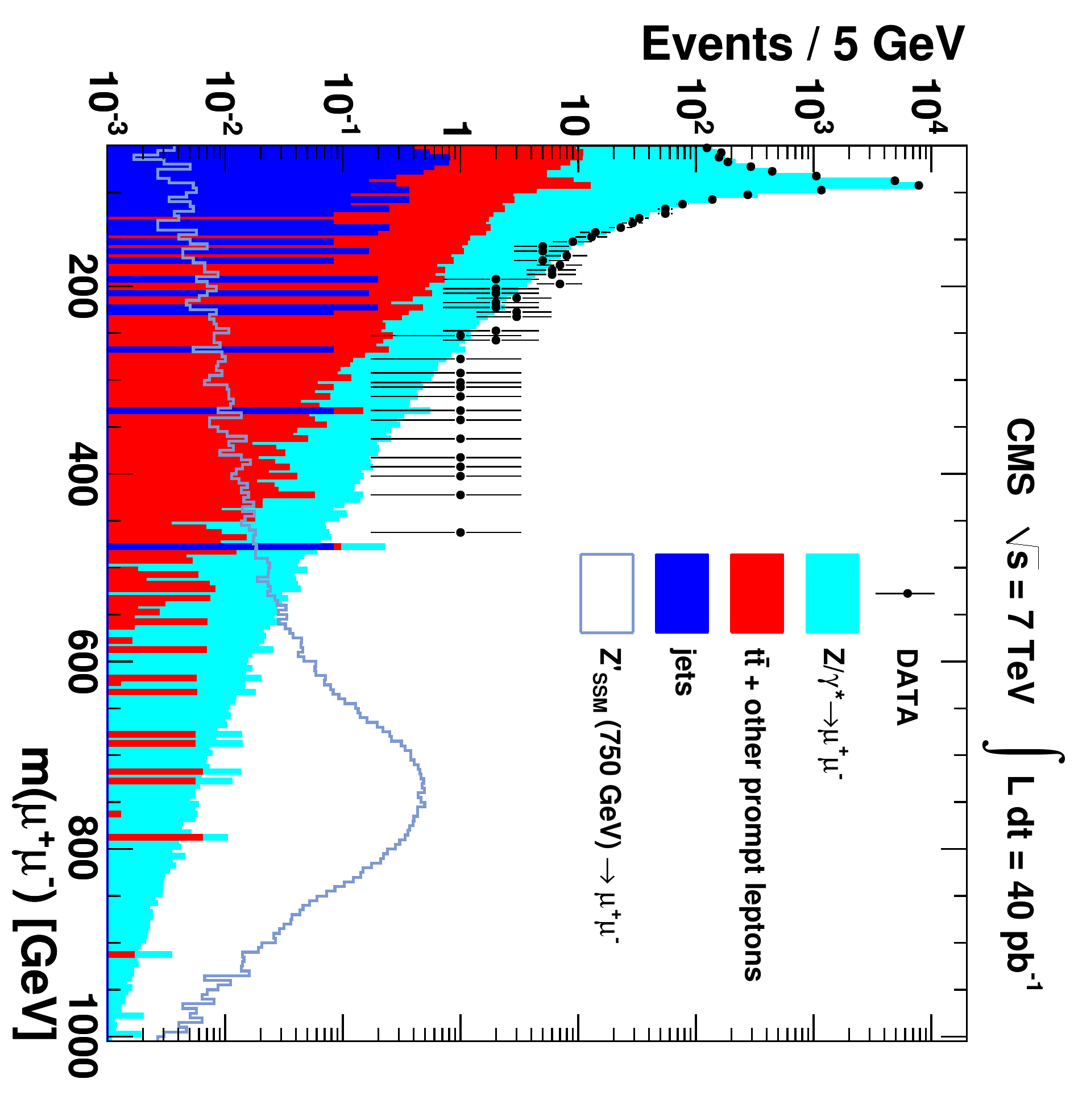}
\includegraphics[angle=90,width=0.49\textwidth]{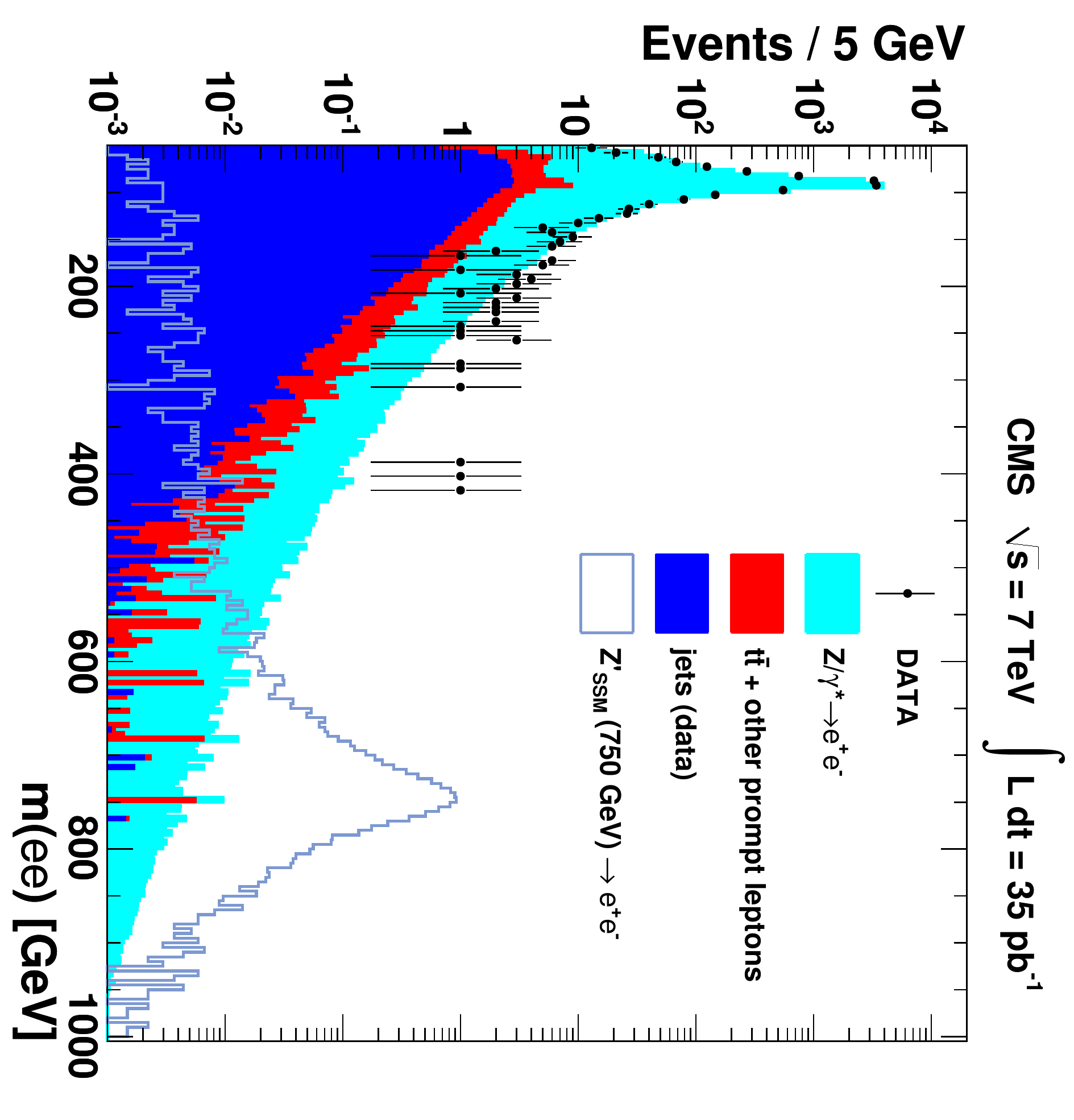}
\end{center}
\caption{\label{fig:spectra}
Invariant mass spectrum of $\Pgmp\Pgmm$ (left) and $\Pe\Pe$ (right) events. The points with
error bars represent the data. The uncertainties on the data points (statistical only) 
represent 68\%  confidence intervals for the Poisson means. The
filled histograms represent the expectations from
SM processes: $\cPZ{/}\gamma^*$, $\ttbar$, other sources of 
prompt leptons ($\tq\PW$, diboson production, $\cPZ\to\Pgt\Pgt$), and
the multi-jet backgrounds.
The open histogram  shows the  signal expected for a $\ZPSSM$ with
a mass of 750\GeV.
}
\end{figure}

The measured dimuon and dielectron invariant mass spectra are displayed in
Figs.~\ref{fig:spectra}(left) and (right) respectively,
along with the expected signal from $\ZPSSM$ with a mass of 750\GeV.
In the dimuon sample, the highest invariant mass
event has $m_{\Pgm\Pgm}=463$\GeV, with the $\pt$ of the two
muons measured to be 258 and 185\GeV.
The highest invariant mass event in the dielectron sample
has $m_{\Pe\Pe}=419$\GeV, with the electron candidates having $\ET$ of 125 and 84\GeV.

\begin{figure*}[htbp!]
\begin{center}
\includegraphics[angle=90,width=0.49\textwidth]{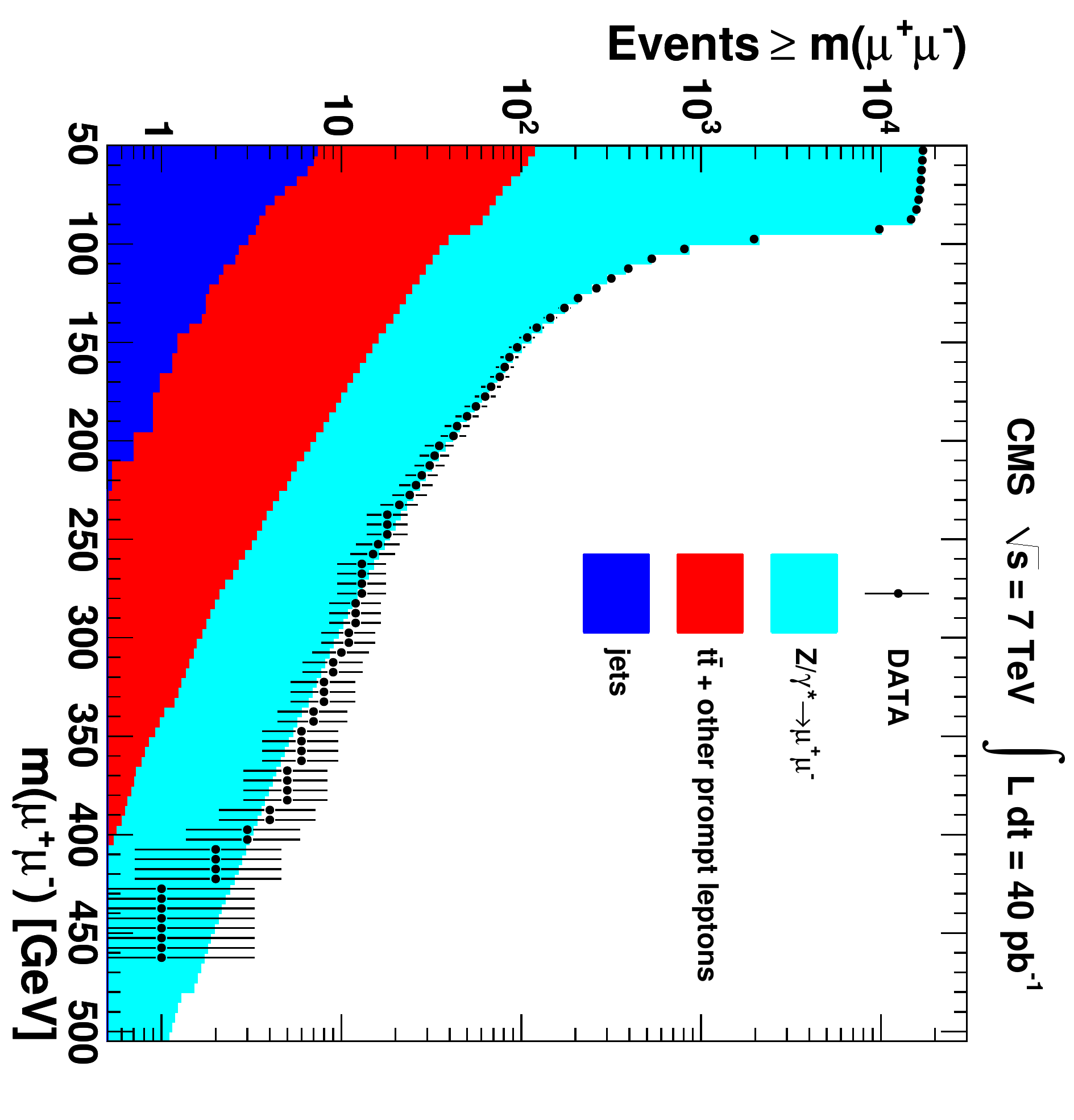}
\includegraphics[angle=90,width=0.49\textwidth]{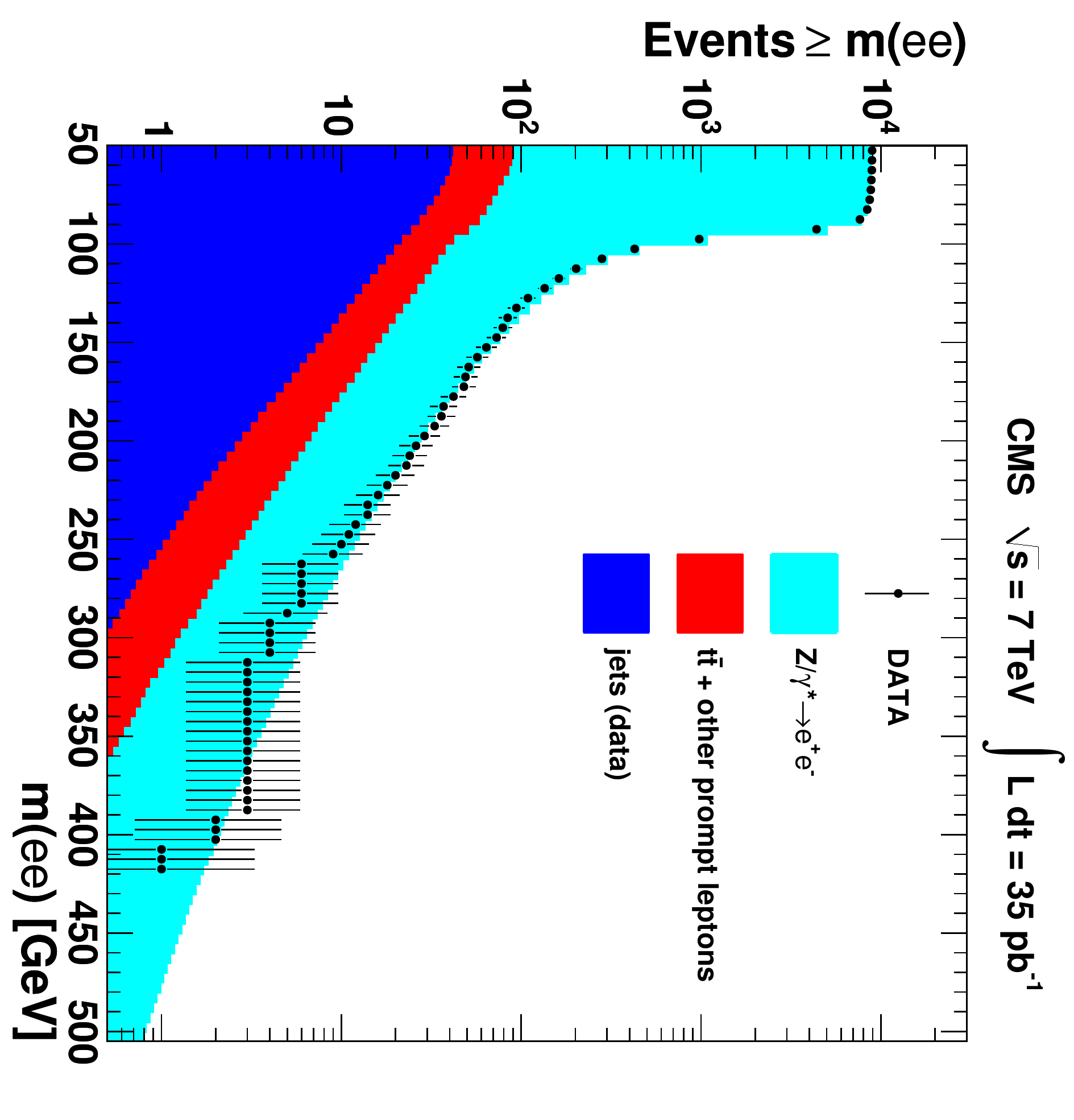}
\end{center}
\caption{\label{fig:cum_spectra} Cumulative distribution of invariant
mass spectrum of $\Pgmp\Pgmm$ (left) and $\Pe\Pe$ (right) events.  The
points with error bars represent the data, and the filled histogram
represents the expectations from SM processes.  }
\end{figure*}

The expectations from the various background sources,
$\cPZ{/}\gamma^*$, $\ttbar$, other sources of  prompt leptons ($\tq\PW$, diboson
production, $\cPZ\to\Pgt\Pgt$) and multi-jet events
are also overlaid in Fig.~\ref{fig:spectra}.  For the dielectron
sample, the multi-jet background estimate was obtained directly from
the data. The prediction for Drell--Yan production of $\cPZ{/}\gamma^*$
is normalized to the observed $\cPZ\to\ell\ell$ signal.  All other MC
predictions are normalized to the expected cross sections.
Figures~\ref{fig:cum_spectra}(left) and (right) show the corresponding
cumulative distributions of the spectra for the dimuon and
dielectron samples.  Good agreement is observed between data and the
expectation from SM processes over the mass region above the Z peak.

Searches for narrow resonances at the Tevatron~\cite{D0_Zp,CDF_Zp}
have placed lower limits in the mass range $600$\GeV to
$1000$\GeV. The region with dilepton masses $120\GeV < m_{\ell\ell} <
200\GeV$ is part of the region for which resonances have been excluded
by previous experiments, and thus should be dominated by SM processes.  The
observed good agreement between the data and the prediction in this
control region gives confidence that the SM expectations and the detector
performance are well understood.

In the $\cPZ$~peak mass region defined as $60 < m_{\ell\ell} <
120$\GeV, the number of dimuon and dielectron candidates are 16\,515
and 8\,768 respectively, with very small backgrounds.  The
difference in the electron and muon numbers is due to the higher $E_T$
cut in the electron analysis and lower electron identification
efficiencies at these energies.  The expected yields in the control
region (120--200\GeV) and high invariant mass regions ($>$ 200\GeV) are listed in
Table~\ref{tab:event_yield}.  The agreement between the observed data
and expectations, while not used in the shape-based analysis,
is good.

\begin{table*}[htb!]
\centering
\caption{Number of dilepton events with invariant mass in the control
region $120~<~m_{\ell\ell}~<~200$\GeV and the search region $m_{\ell\ell} > 200$\GeV.
The expected number of $\cPZpr$ events is given within ranges of 328\GeV and 
120\GeV for the dimuon sample and the dielectron sample respectively, centred 
on 750\GeV. The total background is the sum of the SM processes listed.
The MC yields are normalized to the expected cross sections. 
Uncertainties include both statistical and systematic components
added in quadrature.}
\label{tab:event_yield}
\begin{tabular}{l|c|c|c|c}
\hline\hline
Source & \multicolumn{4}{c}{Number of events} \\
       & \multicolumn{2}{c|}{Dimuon sample }
       & \multicolumn{2}{c}{Dielectron sample} \\
 &  ($120-200$)\GeV  & $ >$200\GeV
 &  ($120-200$)\GeV  & $ >$200\GeV  \\ \hline
 CMS data                   &  227             &  35              & 109             & 26            \\
$\ZPSSM$ (750\GeV)         & ---              & $15.0 \pm 1.9$   & ---             & $8.7\pm1.1$   \\
Total background           & $204 \pm 23$     & $36.3 \pm 4.3$   & $120 \pm 14$    & $24.4\pm 3.0$ \\ \hline
$\cPZ{/}\gamma^*$          & $187 \pm 23$     & $30.2 \pm 3.6$   & $104 \pm 14$    & $18.8\pm 2.3$   \\
$\ttbar$                   & $12.3 \pm 2.3$   & $4.2 \pm 0.8$    & $7.6 \pm 1.4$   & $2.7 \pm 0.5$ \\
Other prompt leptons      & $4.4 \pm 0.5$    & $1.7 \pm 0.2$    & $2.1 \pm 0.2$   & $0.8 \pm 0.1$  \\
Multi-jet events           & $0.6 \pm 0.2$    & $0.2 \pm 0.1$    & $6.5 \pm 2.6$   & $2.1 \pm 0.8$  \\
\hline\hline
\end{tabular}
\end{table*}

\section{Limits on the Production Cross Section}

The observed invariant mass spectrum agrees with expectations based on
standard model processes, therefore limits are set on the possible
contributions from a narrow heavy resonance.
The parameter of interest is the ratio of the products of cross sections
and branching fractions:
\begin{equation}
\label{eq:rsigma}
R_\sigma = \frac{\sigma(\Pp\Pp\to \cPZpr+X\to\ell\ell+X)}
                {\sigma(\Pp\Pp\to \cPZ+X       \to\ell\ell+X)}.
\end{equation}

By focusing on the ratio
$R_\sigma$, we eliminate the uncertainty in the integrated
luminosity, reduce the dependence on experimental
acceptance, trigger, and offline efficiencies, and generally obtain a more robust result.

For statistical inference about $R_\sigma$, we first estimate the Poisson mean
$\mu_\cPZ$ of the number of $\cPZ\to\ell\ell$ events in the sample
by counting the number of events in the $\cPZ$ peak mass region and
correcting for a small ($\sim 0.4\%$) background contamination (determined with MC simulation).
The uncertainty on $\mu_\cPZ$ is about 1\% (almost all statistical) and contributes negligibly to
the uncertainty on $R_\sigma$.

We then construct an extended unbinned likelihood function for the
spectrum of $\ell\ell$ invariant mass values $m$ above
200\GeV, based on a sum of analytic probability density functions
(pdfs) for the signal and background shapes.

The pdf $f_\mathtt{S}(m|\Gamma,M,w)$ for the resonance signal is a
Breit-Wigner of width $\Gamma$ and mass $M$ convoluted with a Gaussian
resolution function of width $w$ (section~\ref{sec:leptonID}). The width $\Gamma$ is taken to
be that of the $\ZPSSM$ (about 3\%); as noted below, the high-mass limits
are insensitive to this width.
The Poisson mean of
the yield is $\mu_\mathtt{S} = R_\sigma \cdot \mu_\cPZ \cdot R_\epsilon$,
where $R_\epsilon$ is the ratio of selection efficiency times detector
acceptance for $\cPZpr$ decay to that of $\cPZ$ decay; $\mu_\mathtt{B}$
denotes the Poisson mean of the total background yield.
A background
pdf $f_\mathtt{B}$ was chosen and its shape parameters fixed by
fitting to the simulated Drell--Yan spectrum in the mass range
$200 < m_{\ell\ell} < 2000$\GeV.
Two functional forms for the dependence of $f_\mathtt{B}$
on shape parameters $\alpha$ and $\kappa$ were tried:
$f_\mathtt{B}(m|\alpha,\kappa) \sim  \exp(-\alpha m^\kappa)$ and
$\sim \exp(-\alpha m)m^{-\kappa}$.  Both
yielded good  fits and consistent results for both the dimuon and dielectron spectra.
For definiteness, this Letter presents results obtained with the latter form.

The extended likelihood ${\cal L}$ is then

\begin{equation}
\label{eq:likelihood}
{\cal L}({\boldsymbol m}|R_\sigma,M,\Gamma,w,\alpha,\kappa,\mu_\mathtt{B}) =
\frac{\mu^N e^{-\mu}}{N!}\prod_{i=1}^{N}\left(
\frac{\mu_\mathtt{S}(R_\sigma)}{\mu}f_\mathtt{S}(m_i|M,\Gamma,w)+
\frac{\mu_\mathtt{B}}{\mu}f_\mathtt{B}(m_i|\alpha,\kappa)
\right),
\end{equation}

where  ${\boldsymbol m}$ denotes the dataset in which
the observables are the invariant mass values of the lepton pairs,
$m_i$; $N$ denotes the total number of events observed above 200\GeV;
and $\mu=\mu_\mathtt{S}+ \mu_\mathtt{B}$ is
the mean of the Poisson distribution from which $N$ is an observation.

Starting from Eqn.~\ref{eq:likelihood}, confidence/credible intervals
are computed using more than one approach, both frequentist (using
profile likelihood ratios) and Bayesian (multiplying ${\cal L}$ by prior
pdfs including a uniform prior for the signal mean).
With no candidate events in the region of small expected background
above 465\GeV, the result is insensitive to
the statistical technique, and also with respect to the width of the $\cPZpr$
and to changes in systematic uncertainties and their functional forms,
taken to be log-normal distributions with fractional uncertainties.

For $R_\epsilon$, we assign an uncertainty of 8\% for the dielectron
channel and 3\% for the dimuon channel. These values reflect our current
understanding of the detector acceptance and reconstruction efficiency
turn-on at low mass (including PDF uncertainties
and mass-dependence of $k$-factors), as well as the corresponding values
at high mass, where
cosmic ray muons are available to study muon performance but not electron
performance.  The uncertainty in the mass scale affects only the mass
region below 500\GeV where there are events in both channels
extrapolating from the well-calibrated observed resonances.
For the dielectron channel, it is set to 1\% based on linearity studies.
For the dimuon channel, it is set to zero, as a sensitivity study showed
negligible change in the results up to the maximum misalignment consistent
with alignment studies (corresponding to several percent change in momentum scale).
The acceptance for $\GKK$ (spin 2) is higher than for $\cPZpr$ (spin 1) by
less than 8\% over the mass range 0.75--1.1 TeV. This was
conservatively neglected when calculating the limits.

In the frequentist calculation, the mean background level
$\mu_\mathtt{B}$
is the maximum likelihood estimate; in the fully
 Bayesian
calculation a prior must be assigned to the mean background
 level,
but the result is insensitive to reasonable choices (i.e., for which
the likelihood dominates the prior).

The upper limits on $R_{\sigma}$ (Eqn.~\ref{eq:rsigma}) from the various approaches
are similar, and we report the Bayesian
result (implemented with Markov Chain Monte Carlo in
{\sc RooStats}~\cite{MCMC}) for definiteness.
From the dimuon and dielectron data, we obtain the upper limits on the cross section
ratio $R_{\sigma}$ at 95\% confidence level (C.L.) shown in
Figs.~\ref{fig:limits}(upper) and (middle), respectively.

In Fig.~\ref{fig:limits}, the predicted cross section ratios
for $\ZPSSM$ and $\ZPPSI$  production are superimposed
together with those
for $\GKK$ production with dimensionless graviton coupling
to SM fields $k/\overline{M}_\mathrm{Pl}=0.05$ and $0.1$.
The leading order cross section predictions for
$\ZPSSM$ and $\ZPPSI$
from \PYTHIA using {\sc CTEQ6.1} PDFs are corrected for a mass dependent
$k$-factor obtained using {\textsc ZWPRODP}~\cite{Accomando:2010fz,Hamberg:1990np,
vanNeerven:1991gh,ZWPROD} to account for NNLO contributions.
For the RS graviton model, a constant NLO $k$-factor of 1.6 is
used~\cite{Mathews:2005bw}. The uncertainties
due to the QCD scale parameter and PDFs are indicated as a band.
The NNLO prediction for the $\cPZ$ production cross
section is 0.97$\pm\,$0.04~nb~\cite{Melnikov:2006kv}.

Propagating the above-mentioned uncertainties into the
comparison of the experimental limits with the predicted cross section
ratios, we exclude
at 95\%~C.L. $\cPZpr$ masses as follows.  From the dimuon only
analysis,
the $\ZPSSM$ can be excluded below 1027\GeV, the
$\ZPPSI$ below 792\GeV, and the
RS $\GKK$ below 778 (987)\GeV for couplings
of 0.05 (0.1). For the dielectron analysis, the
production of $\ZPSSM$ and $\ZPPSI$ bosons is
excluded for masses below 958 and 731\GeV, respectively. The
corresponding lower limits on the mass for
RS $\GKK$ with
couplings of 0.05 (0.10) are 729 (931)\GeV.

\subsection{Combined Limits on the Production Cross Section Using
Dimuon and Dielectron Events}

\begin{figure}[htbp!]
\begin{center}
\includegraphics[width=0.65\textwidth,angle=0]{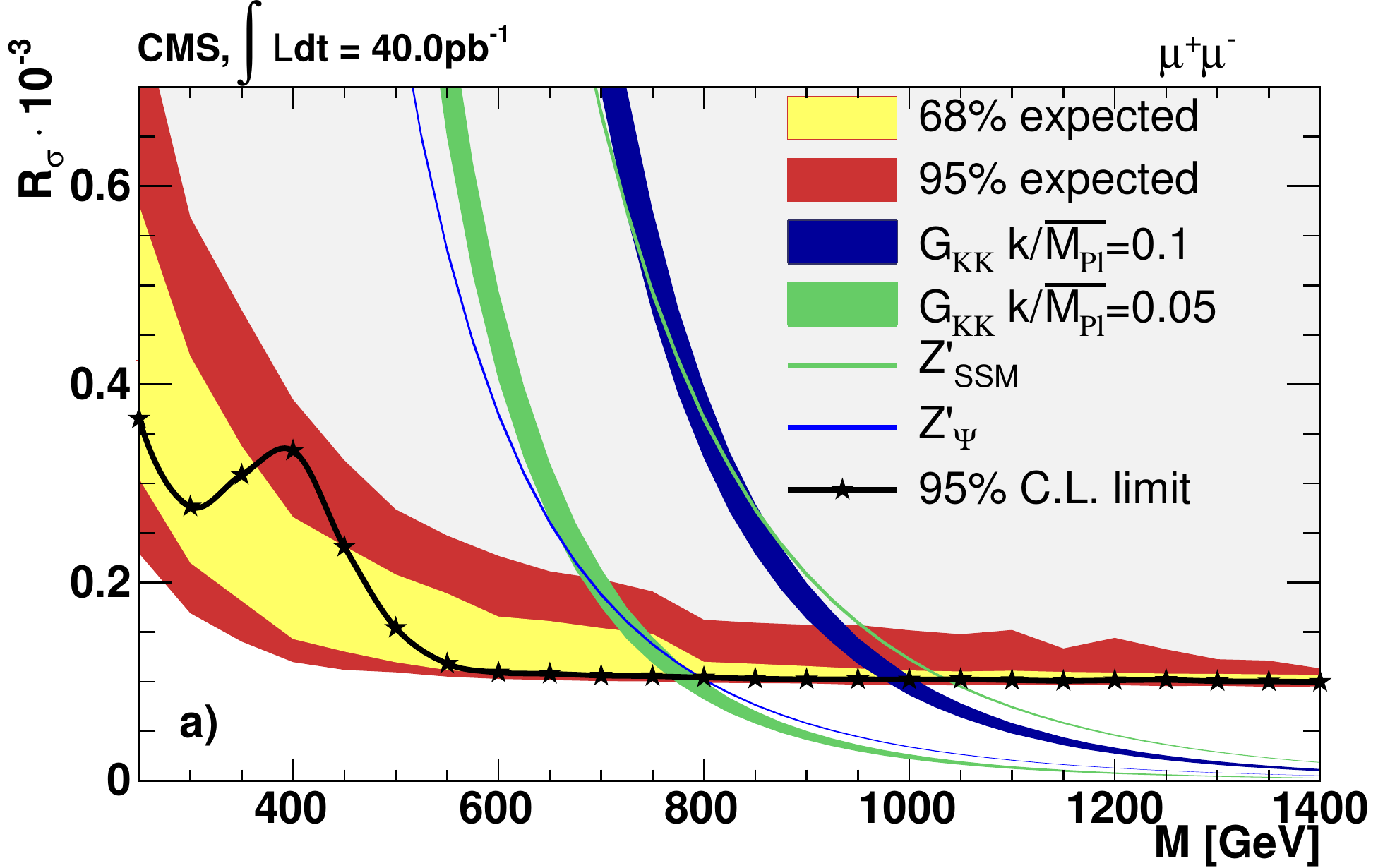}
\includegraphics[width=0.65\textwidth,angle=0]{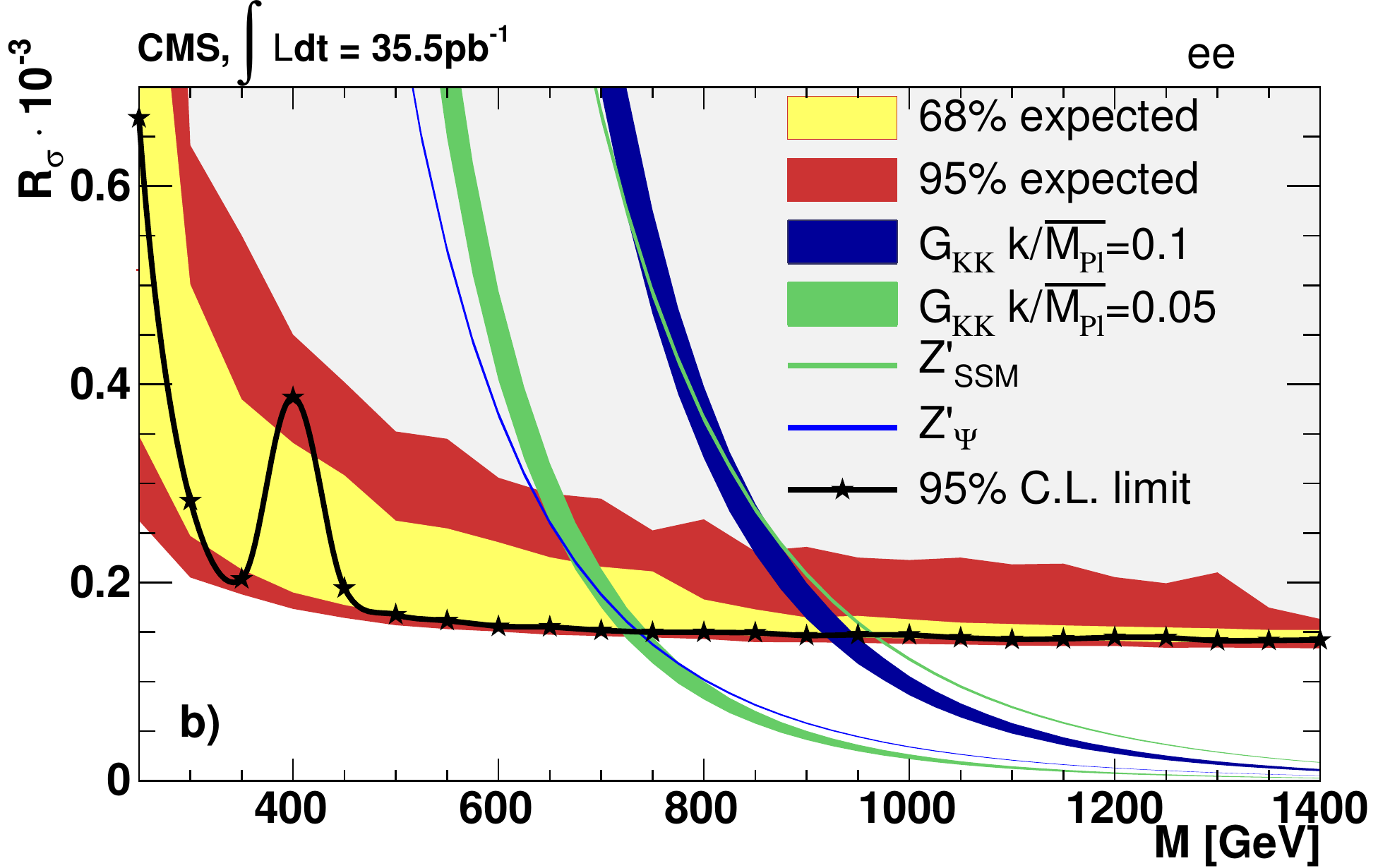}
\includegraphics[width=0.65\textwidth,angle=0]{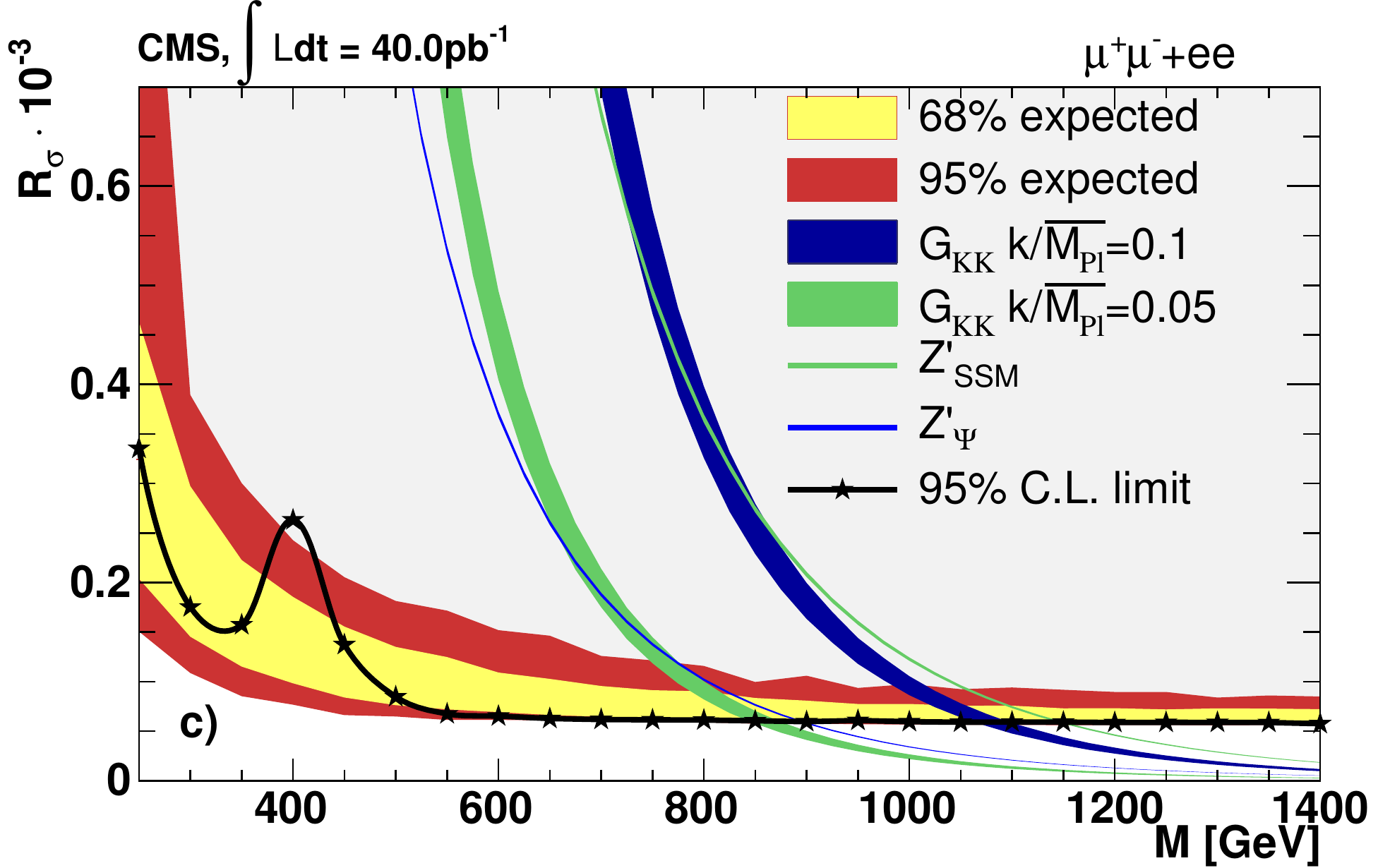}
\end{center}
\caption{\label{fig:limits} Upper limits as a function
of resonance mass $M$,
on the production ratio $R_{\sigma}$ of
 cross section times branching fraction into lepton pairs
  for $\ZPSSM$ and $\GKK$ production and $\ZPPSI$
 boson production. The limits are shown from (top) the $\Pgmp\Pgmm$ final
 state, (middle) the $\Pe\Pe$ final state and (bottom) the combined dilepton result.
Shaded yellow and red  bands correspond to the $68\%$ and  $95\%$ quantiles 
for the expected  limits.
The predicted cross section ratios are shown as bands, with widths
indicating the theoretical uncertainties. 
}

\end{figure}

The above statistical formalism is generalized to combine the results from the
dimuon and dielectron channels,
by defining the combined likelihood as the product
of the likelihoods for the individual channels with $R_\sigma$ forced to
be the same value for both channels. The combined limit is shown in
Fig.~\ref{fig:limits}~(bottom).

By combining the two channels, the following
 95\% C.L. lower limits on the mass of a $\cPZpr$ resonance are obtained:
1140\GeV for the $\ZPSSM$, and 887\GeV for $\ZPPSI$ models. RS
Kaluza--Klein gravitons are excluded below 855 (1079)\GeV
 for values of couplings 0.05 (0.10).
Our observed limits are more restrictive than or comparable to
those previously obtained via similar direct searches by the
Tevatron experiments~\cite{D0_RS,D0_Zp,CDF_RS,CDF_Zp},
or indirect searches by LEP-II experiments~\cite{delphi,aleph,opal,l3},
with the exception of $\ZPSSM$, where the value from LEP-II
is the most restrictive.

The distortion of the observed limits at $\sim$400\GeV visible in
Fig.~\ref{fig:limits} is the result of a clustering of several dimuon
and dielectron events around this mass.  We have tested for the
statistical significance of these excesses ($p$-values expressed as
equivalent $Z$-values, i.e. effective number of Gaussian sigma in a
one-sided test), using the techniques described in \cite{PTDR2}.
For the dimuon sample, the probability of an enhancement at
least as large as that at 400 GeV occurring anywhere above 200 GeV in
the observed sample size corresponds to $Z<0.2$; for the electron
sample, it is less. For the combined data sample, the corresponding
probability in a joint peak search is equivalent to $Z=1.1$.

In the narrow-width approximation, the cross section for the process
$\Pp\Pp\to \cPZpr+X\to\ell\ell+X$ can be
expressed~\cite{Carena:2004xs,Accomando:2010fz} in terms of the
quantity $c_u w_u + c_d w_d$, where $c_u$ and $c_d$ contain the
information from the model-dependent $\cPZpr$ couplings to fermions
in the annihilation of charge 2/3 and charge $-$1/3 quarks,
respectively, and where $w_u$ and $w_d$ contain the information about
PDFs for the respective annihilation at a given $\cPZpr$ mass.

\begin{figure}[htbp!]
\begin{center}
\includegraphics[width=0.6\textwidth]{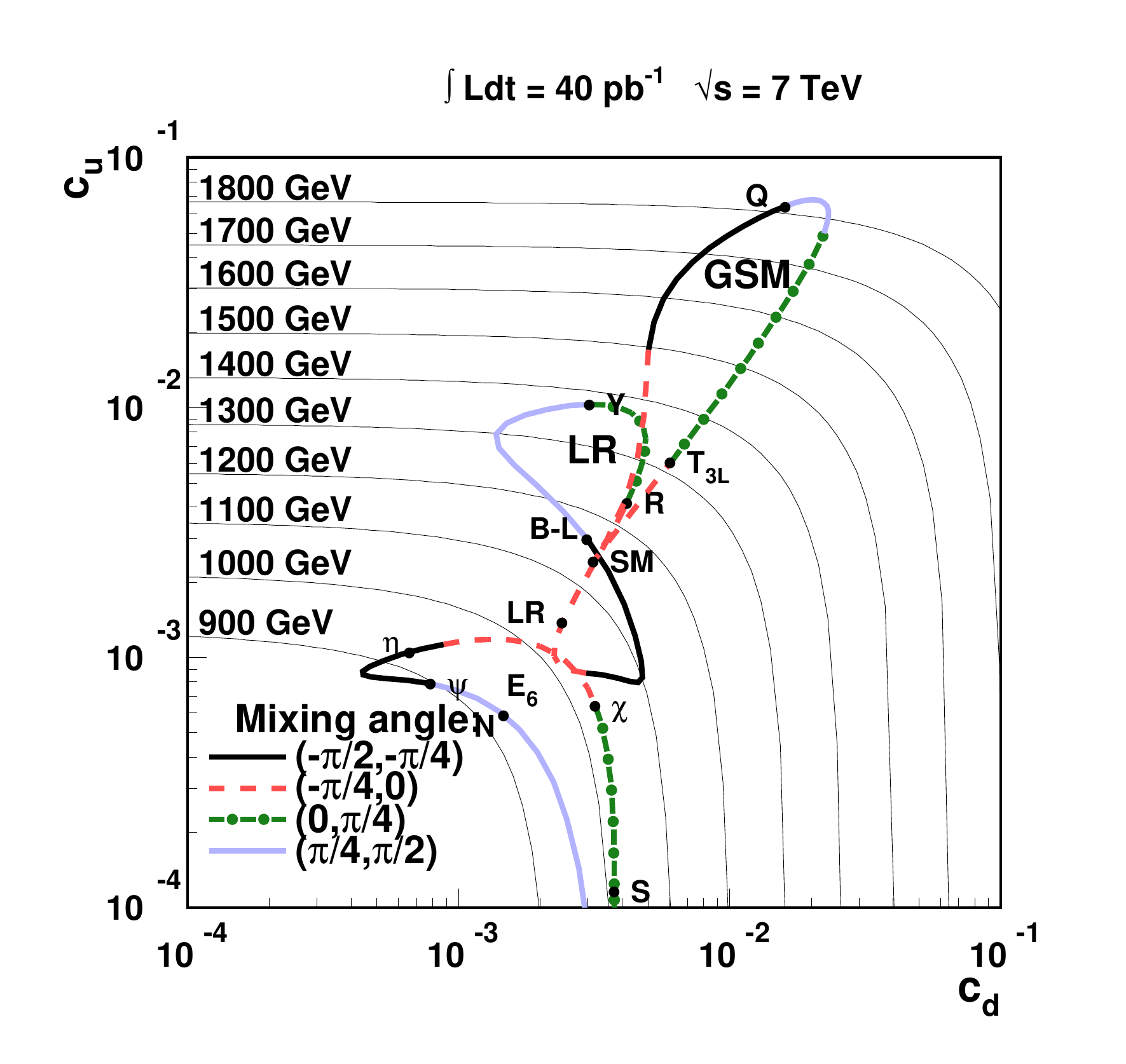}
\end{center}
\caption{\label{fig:CuCd}
95\% C.L. lower limits on the $\cPZpr$ mass, represented by the thin continuous lines
in the $(c_d,c_u)$
plane.  Curves for three classes of model are shown.  Colours on the
curves correspond to different mixing angles of the generators defined
in each model. For any point on a curve, the mass limit corresponding
to that value of $(c_d,c_u)$ is given by the intersected contour.}
\end{figure}

The
translation of the experimental limits into the ($c_u$,$c_d$) plane has
been studied in the context of both the narrow-width and finite width
approximations. The procedures have been shown to give the same
results. In Fig.~\ref{fig:CuCd} the limits on the $\cPZpr$ mass are
shown as lines in the $(c_d,c_u)$ plane intersected by curves from
various models which specify $(c_d,c_u)$ as a function of a model
mixing parameter.
In this plane, the thin solid lines labeled by mass are
iso-contours of cross section with constant
$c_u + (w_d/w_u)c_d$, where $w_d/w_u$ is in the range 0.5--0.6 for the
results relevant here.  As this linear combination
increases or decreases by an order of magnitude, the mass limits
change by roughly 500 GeV. 
The point labeled SM corresponds to the
$\ZPSSM$; it lies on the more general curve for the
Generalized Sequential Standard Model (GSM) for which the generators
of the $U(1)_{T_{3L}}$ and $U(1)_Q$ gauge groups are mixed with a mixing
angle $\alpha$. Then $\alpha = -0.072\pi$ corresponds to the
$Z^\prime_{SSM}$ and $\alpha=0$ and $\pi/2$ define the $T_{3L}$ and
$Q$ benchmarks, respectively, which have larger values of $(c_d,c_u)$
and hence larger lower bounds on the masses.  Also shown are contours
for the E$_6$ model (with $\chi$, $\psi$, $\eta$, $S$, and $N$
corresponding to angles 0, 0.5$\pi$, $-0.29\pi$, 0.13$\pi$, and
0.42$\pi$, respectively) and Generalized LR models (with $R$, $B-L$,
$LR$, and $Y$ corresponding to angles 0, 0.5$\pi$, $-0.13\pi$, and
0.25$\pi$, respectively)~\cite{Accomando:2010fz} .

\section{Summary}

The CMS Collaboration has searched for narrow resonances in the
invariant mass spectrum of  dimuon and dielectron final
states in event samples corresponding to an integrated luminosity of
$40$\pbinv and $35$\pbinv, respectively. The spectra are consistent
with standard model expectations and upper limits on the
cross section times branching fraction for $\cPZpr$ into
lepton pairs relative to standard model $\cPZ$ boson
production have been set.
Mass limits have been set on neutral gauge bosons $\cPZpr$
and RS Kaluza--Klein gravitons $\GKK$.  A $\cPZpr$ with
standard-model-like couplings can be excluded below 1140\GeV, the
superstring-inspired $\ZPPSI$ below 887\GeV, and RS
Kaluza--Klein gravitons below 855 (1079)\GeV for couplings of
0.05 (0.10), all at 95\%~C.L.
The higher centre of mass energy used in this search, compared to that
of previous experiments, has resulted in limits that are comparable to or
exceed those previously published, despite the much lower
integrated luminosity accumulated at the LHC thus far.

\section*{Acknowledgments}

We wish to congratulate our colleagues in the CERN accelerator
departments for the excellent performance of the LHC machine. We thank
the technical and administrative staff at CERN and other CMS
institutes, and acknowledge support from: FMSR (Austria); FNRS and FWO
(Belgium); CNPq, CAPES, FAPERJ, and FAPESP (Brazil); MES (Bulgaria);
CERN; CAS, MoST, and NSFC (China); COLCIENCIAS (Colombia); MSES
(Croatia); RPF (Cyprus); Academy of Sciences and NICPB (Estonia);
Academy of Finland, ME, and HIP (Finland); CEA and CNRS/IN2P3
(France); BMBF, DFG, and HGF (Germany); GSRT (Greece); OTKA and NKTH
(Hungary); DAE and DST (India); IPM (Iran); SFI (Ireland); INFN
(Italy); NRF and WCU (Korea); LAS (Lithuania); CINVESTAV, CONACYT,
SEP, and UASLP-FAI (Mexico); PAEC (Pakistan); SCSR (Poland); FCT
(Portugal); JINR (Armenia, Belarus, Georgia, Ukraine, Uzbekistan); MST
and MAE (Russia); MSTD (Serbia); MICINN and CPAN (Spain); Swiss
Funding Agencies (Switzerland); NSC (Taipei); TUBITAK and TAEK
(Turkey); STFC (United Kingdom); DOE and NSF (USA).

\bibliography{auto_generated}   

\providecommand{\href}[2]{#2}\begingroup\raggedright\begin{thebibliography}{10}%
\makeatletter
\providecommand{\hrefCMSnoop }[0]{\@secondoftwo}%
\makeatother

\bibitem{lhc}
\hrefCMSnoop {} {L.~Evans and P.~Bryant~(editors), ``{LHC Machine}'',} \textit{
  JINST} \textbf{ 3} (2008) S08001.
\href{http://dx.doi.org/10.1088/1748-0221/3/08/S08001}{\texttt{
  doi:10.1088/1748-0221/3/08/S08001}}.

\bibitem{Leike:1998wr}
\hrefCMSnoop {} {A.~Leike, ``{The phenomenology of extra neutral gauge
  bosons}'',} \textit{ Phys. Rept.} \textbf{ 317} (1999) 143,
  \href{http://www.arXiv.org/abs/hep-ph/9805494}{\texttt{
  arXiv:hep-ph/9805494}}.
\href{http://dx.doi.org/10.1016/S0370-1573(98)00133-1}{\texttt{
  doi:10.1016/S0370-1573(98)00133-1}}.

\bibitem{Randall:1999vf}
\hrefCMSnoop {} {L.~Randall and R.~Sundrum, ``{An alternative to
  compactification}'',} \textit{ Phys. Rev. Lett.} \textbf{ 83} (1999) 4690,
  \href{http://www.arXiv.org/abs/hep-th/9906064}{\texttt{
  arXiv:hep-th/9906064}}.
\href{http://dx.doi.org/10.1103/PhysRevLett.83.4690}{\texttt{
  doi:10.1103/PhysRevLett.83.4690}}.

\bibitem{Randall:1999ee}
\hrefCMSnoop {} {L.~Randall and R.~Sundrum, ``A large mass hierarchy from a
  small extra dimension'',} \textit{ Phys. Rev. Lett.} \textbf{ 83} (1999)
  3370, \href{http://www.arXiv.org/abs/hep-ph/9905221}{\texttt{
  arXiv:hep-ph/9905221}}.
\href{http://dx.doi.org/10.1103/PhysRevLett.83.3370}{\texttt{
  doi:10.1103/PhysRevLett.83.3370}}.

\bibitem{D0_RS}
\hrefCMSnoop {} {{ D0} Collaboration, ``Search for {Randall-Sundrum} gravitons
  in the dielectron and diphoton final states with 5.4~fb$^{-1}$ of data from
  ${p\bar{p}}$ collisions at $\sqrt{s} = 1.96$~{TeV}'',} \textit{ Phys. Rev.
  Lett.} \textbf{ 104} (2010) 241802,
  \href{http://www.arXiv.org/abs/1004.1826}{\texttt{ arXiv:1004.1826}}.
\href{http://dx.doi.org/10.1103/PhysRevLett.104.241802}{\texttt{
  doi:10.1103/PhysRevLett.104.241802}}.

\bibitem{D0_Zp}
\hrefCMSnoop {} {{ D0} Collaboration, ``{Search for a heavy neutral gauge boson
  in the dielectron channel with 5.4~fb$^{-1}$ of ${p\bar{p}}$ collisions at
  ${\sqrt{s} = 1.96}$~TeV}'',} \textit{ Phys. Lett. B} \textbf{ 695} (2011)
  088, \href{http://www.arXiv.org/abs/1008.2023}{\texttt{ arXiv:1008.2023}}.
\href{http://dx.doi.org/10.1016/j.physletb.2010.10.059}{\texttt{
  doi:10.1016/j.physletb.2010.10.059}}.

\bibitem{CDF_RS}
\hrefCMSnoop {} {{ CDF} Collaboration, ``{A search for high-mass resonances
  decaying to dimuons at CDF}'',} \textit{ Phys. Rev. Lett.} \textbf{ 102}
  (2009) 091805, \href{http://www.arXiv.org/abs/0811.0053}{\texttt{
  arXiv:0811.0053}}.
\href{http://dx.doi.org/10.1103/PhysRevLett.102.091805}{\texttt{
  doi:10.1103/PhysRevLett.102.091805}}.

\bibitem{CDF_Zp}
\hrefCMSnoop {} {{ CDF} Collaboration, ``{Search for High-Mass $e^+e^-$
  Resonances in $p\bar{p}$ Collisions at $\sqrt{s} = 1.96$~TeV}'',} \textit{
  Phys. Rev. Lett.} \textbf{ 102} (2009) 031801,
  \href{http://www.arXiv.org/abs/0810.2059}{\texttt{ arXiv:0810.2059}}.
\href{http://dx.doi.org/10.1103/PhysRevLett.102.031801}{\texttt{
  doi:10.1103/PhysRevLett.102.031801}}.

\bibitem{delphi}
\hrefCMSnoop {} {{ DELPHI} Collaboration, ``{Measurement and interpretation of
  fermion-pair production at LEP energies above the Z resonance}'',} \textit{
  Eur. Phys. J.} \textbf{ C45} (2006) 589,
  \href{http://www.arXiv.org/abs/hep-ex/0512012}{\texttt{
  arXiv:hep-ex/0512012}}.
\href{http://dx.doi.org/10.1140/epjc/s2005-02461-0}{\texttt{
  doi:10.1140/epjc/s2005-02461-0}}.

\bibitem{aleph}
\hrefCMSnoop {} {{ ALEPH} Collaboration, ``{Fermion pair production in $e^{+}
  e^{-}$ collisions at 189-209-GeV and constraints on physics beyond the
  standard model}'',} \textit{ Eur. Phys. J.} \textbf{ C49} (2007) 411,
  \href{http://www.arXiv.org/abs/hep-ex/0609051}{\texttt{
  arXiv:hep-ex/0609051}}.
\href{http://dx.doi.org/10.1140/epjc/s10052-006-0156-8}{\texttt{
  doi:10.1140/epjc/s10052-006-0156-8}}.

\bibitem{opal}
\hrefCMSnoop {} {{ OPAL} Collaboration, ``{Tests of the standard model and
  constraints on new physics from measurements of fermion pair production at
  189-GeV to 209-GeV at LEP}'',} \textit{ Eur. Phys. J.} \textbf{ C33} (2004)
  173, \href{http://www.arXiv.org/abs/hep-ex/0309053}{\texttt{
  arXiv:hep-ex/0309053}}.
\href{http://dx.doi.org/10.1140/epjc/s2004-01595-9}{\texttt{
  doi:10.1140/epjc/s2004-01595-9}}.

\bibitem{l3}
\hrefCMSnoop {} {{ L3} Collaboration, ``{Measurement of hadron and lepton-pair
  production in e+ e- collisions at $\sqrt{s}$ = 192-GeV - 208-GeV at LEP}'',}
  \textit{ Eur. Phys. J.} \textbf{ C47} (2006) 1,
  \href{http://www.arXiv.org/abs/hep-ex/0603022}{\texttt{
  arXiv:hep-ex/0603022}}.
\href{http://dx.doi.org/10.1140/epjc/s2006-02539-1}{\texttt{
  doi:10.1140/epjc/s2006-02539-1}}.

\bibitem{Carena:2004xs}
\hrefCMSnoop {} {M.~Carena {et~al.}, ``{$Z^\prime$ gauge bosons at the
  Tevatron}'',} \textit{ Phys. Rev.} \textbf{ D70} (2004) 093009,
  \href{http://www.arXiv.org/abs/hep-ph/0408098}{\texttt{
  arXiv:hep-ph/0408098}}.
\href{http://dx.doi.org/10.1103/PhysRevD.70.093009}{\texttt{
  doi:10.1103/PhysRevD.70.093009}}.

\bibitem{JINST}
\hrefCMSnoop {} {{ CMS} Collaboration, ``{The CMS experiment at the CERN
  LHC}'',} \textit{ JINST} \textbf{ 3} (2008) S08004.
\href{http://dx.doi.org/10.1088/1748-0221/3/08/S08004}{\texttt{
  doi:10.1088/1748-0221/3/08/S08004}}.

\bibitem{EWK-10-002-PAS}
\hrefCMSnoop {} {{ CMS} Collaboration, ``Measurements of Inclusive {$W$} and
  {$Z$} Cross Sections in pp collisions at $\sqrt{s} = 7$ {TeV}'',} \textit{
  JHEP} \textbf{ 01} (2011) 080,
  \href{http://www.arXiv.org/abs/1012.2466}{\texttt{ arXiv:1012.2466}}.
\href{http://dx.doi.org/10.1007/JHEP01(2011)080}{\texttt{
  doi:10.1007/JHEP01(2011)080}}.

\bibitem{MUONPAS}
\href {http://cdsweb.cern.ch/record/1279140} {{ CMS} Collaboration,
  ``Performance of {CMS} muon identification in pp collisions at $\sqrt{s} = 7$
  {TeV}'',} \textit{ CMS Physics Analysis Summary} \textbf{ MUO-10-002}.

\bibitem{EGMPAS}
\href {http://cdsweb.cern.ch/record/1299116} {{ CMS} Collaboration, ``Electron
  reconstruction and identification at $\sqrt(s) = 7$ {TeV}'',} \textit{ CMS
  Physics Analysis Summary} \textbf{ EGM-10-004} (2010).

\bibitem{CMS_CFT_09_014}
\hrefCMSnoop {} {{ CMS} Collaboration, ``Performance of {CMS} Muon
  Reconstruction in Cosmic-Ray Events'',} \textit{ JINST} \textbf{ 5} (2010)
  T03022, \href{http://www.arXiv.org/abs/0911.4994}{\texttt{ arXiv:0911.4994}}.
\href{http://dx.doi.org/10.1088/1748-0221/5/03/T03022}{\texttt{
  doi:10.1088/1748-0221/5/03/T03022}}.

\bibitem{Sjostrand:2006za}
\hrefCMSnoop {} {T.~{\protect Sj\"{o}strand}, S.~Mrenna, and P.~Z. Skands,
  ``{PYTHIA 6.4 Physics and Manual}'',} \textit{ JHEP} \textbf{ 05} (2006) 026,
  \href{http://www.arXiv.org/abs/hep-ph/0603175}{\texttt{
  arXiv:hep-ph/0603175}}.
\href{http://dx.doi.org/10.1088/1126-6708/2006/05/026}{\texttt{
  doi:10.1088/1126-6708/2006/05/026}}.

\bibitem{MADGRAPH}
\hrefCMSnoop {} {F.~Maltoni and T.~Stelzer, ``{MadEvent}: Automatic event
  generation with {MadGraph}'',} \textit{ JHEP} \textbf{ 02} (2003) 027,
  \href{http://www.arXiv.org/abs/hep-ph/0208156v1}{\texttt{
  arXiv:hep-ph/0208156v1}}.
  \href{http://dx.doi.org/10.1088/1126-6708/2003/02/027}{\texttt{
  doi:10.1088/1126-6708/2003/02/027}}.

\bibitem{Alioli:2008gx}
\hrefCMSnoop {} {S.~Alioli {et~al.}, ``{NLO vector-boson production matched
  with shower in POWHEG}'',} \textit{ JHEP} \textbf{ 07} (2008) 060,
  \href{http://www.arXiv.org/abs/0805.4802}{\texttt{ arXiv:0805.4802}}.
\href{http://dx.doi.org/10.1088/1126-6708/2008/07/060}{\texttt{
  doi:10.1088/1126-6708/2008/07/060}}.

\bibitem{Nason:2004rx}
\hrefCMSnoop {} {P.~Nason, ``{A new method for combining NLO QCD with shower
  Monte Carlo algorithms}'',} \textit{ JHEP} \textbf{ 11} (2004) 040,
  \href{http://www.arXiv.org/abs/hep-ph/0409146}{\texttt{
  arXiv:hep-ph/0409146}}.
\href{http://dx.doi.org/10.1088/1126-6708/2004/11/040}{\texttt{
  doi:10.1088/1126-6708/2004/11/040}}.

\bibitem{Frixione:2007vw}
\hrefCMSnoop {} {S.~Frixione, P.~Nason, and C.~Oleari, ``{Matching NLO QCD
  computations with Parton Shower simulations: the POWHEG method}'',} \textit{
  JHEP} \textbf{ 11} (2007) 070,
  \href{http://www.arXiv.org/abs/0709.2092}{\texttt{ arXiv:0709.2092}}.
\href{http://dx.doi.org/10.1088/1126-6708/2007/11/070}{\texttt{
  doi:10.1088/1126-6708/2007/11/070}}.

\bibitem{Pumplin:2002vw}
\hrefCMSnoop {} {J.~Pumplin {et~al.}, ``{New generation of parton distributions
  with uncertainties from global QCD analysis}'',} \textit{ JHEP} \textbf{ 07}
  (2002) 012, \href{http://www.arXiv.org/abs/hep-ph/0201195}{\texttt{
  arXiv:hep-ph/0201195}}.
\href{http://dx.doi.org/10.1088/1126-6708/2002/07/012}{\texttt{
  doi:10.1088/1126-6708/2002/07/012}}.

\bibitem{GEANT4}
\hrefCMSnoop {} {{ GEANT4} Collaboration, ``{GEANT4: A simulation toolkit}'',}
  \textit{ Nucl. Instrum. Meth.} \textbf{ A506} (2003) 250.
\href{http://dx.doi.org/10.1016/S0168-9002(03)01368-8}{\texttt{
  doi:10.1016/S0168-9002(03)01368-8}}.

\bibitem{Melnikov:2006kv}
\hrefCMSnoop {} {K.~Melnikov and F.~Petriello, ``{Electroweak gauge boson
  production at hadron colliders through ${\cal O}(\alpha_s^2)$}'',} \textit{
  Phys. Rev.} \textbf{ D74} (2006) 114017,
  \href{http://www.arXiv.org/abs/hep-ph/0609070}{\texttt{
  arXiv:hep-ph/0609070}}.
\href{http://dx.doi.org/10.1103/PhysRevD.74.114017}{\texttt{
  doi:10.1103/PhysRevD.74.114017}}.

\bibitem{Stump:2003yu}
\hrefCMSnoop {} {D.~Stump {et~al.}, ``{Inclusive jet production, parton
  distributions, and the search for new physics}'',} \textit{ JHEP} \textbf{
  10} (2003) 046, \href{http://www.arXiv.org/abs/hep-ph/0303013}{\texttt{
  arXiv:hep-ph/0303013}}.
\href{http://dx.doi.org/10.1088/1126-6708/2003/10/046}{\texttt{
  doi:10.1088/1126-6708/2003/10/046}}.

\bibitem{Bourilkov:2003kk}
\hrefCMSnoop {} {D.~Bourilkov, ``{Study of parton density function
  uncertainties with LHAPDF and PYTHIA at LHC}'',} (2003).
  \href{http://www.arXiv.org/abs/hep-ph/0305126}{\texttt{
  arXiv:hep-ph/0305126}}.


\bibitem{Whalley:2005nh}
\hrefCMSnoop {} {M.~R. Whalley, D.~Bourilkov, and R.~C. Group, ``{The Les
  Houches Accord PDFs (LHAPDF) and Lhaglue}'',} (2005).
  \href{http://www.arXiv.org/abs/hep-ph/0508110}{\texttt{
  arXiv:hep-ph/0508110}}.


\bibitem{Martin:2007bv}
\hrefCMSnoop {} {A.~D. Martin {et~al.}, ``{Update of parton distributions at
  NNLO}'',} \textit{ Phys. Lett.} \textbf{ B652} (2007) 292,
  \href{http://www.arXiv.org/abs/0706.0459}{\texttt{ arXiv:0706.0459}}.
\href{http://dx.doi.org/10.1016/j.physletb.2007.07.040}{\texttt{
  doi:10.1016/j.physletb.2007.07.040}}.

\bibitem{Campbell:2010ff}
\hrefCMSnoop {} {J.~M. Campbell and R.~Ellis, ``{MCFM for the Tevatron and the
  LHC}'',} \textit{ Nucl. Phys. B (Proc. Suppl.)} \textbf{ 205-206} (2010) 10,
  \href{http://www.arXiv.org/abs/1007.3492}{\texttt{ arXiv:1007.3492}}.
  \href{http://dx.doi.org/10.1016/j.nuclphysbps.2010.08.011}{\texttt{
  doi:10.1016/j.nuclphysbps.2010.08.011}}.

\bibitem{Kleiss:1988xr}
\hrefCMSnoop {} {R.~Kleiss and W.~Stirling, ``{Top quark production at hadron
  colliders: Some useful formulae}'',} \textit{ Z. Phys.} \textbf{ C40} (1988)
  419. \href{http://dx.doi.org/10.1007/BF01548856}{\texttt{
  doi:10.1007/BF01548856}}.

\bibitem{MCMC}
\hrefCMSnoop {} {L.~Moneta {et~al.}, ``{The RooStats project}'',}.
  \href{http://www.arXiv.org/abs/1009.1003}{\texttt{ arXiv:1009.1003}}.

\bibitem{Accomando:2010fz}
\hrefCMSnoop {} {E.~Accomando {et~al.}, ``{Z' physics with early LHC data}'',}
  (2010). \href{http://www.arXiv.org/abs/1010.6058}{\texttt{ arXiv:1010.6058}}.


\bibitem{Hamberg:1990np}
\hrefCMSnoop {} {R.~Hamberg, W.~L. van Neerven, and T.~Matsuura, ``{A complete
  calculation of the order $\alpha_S^{2}$ correction to the Drell--Yan $K$
  factor}'',} \textit{ Nucl. Phys.} \textbf{ B359} (1991) 343. Erratum-ibid.
  {\bf B644} (2002) 403.
\href{http://dx.doi.org/10.1016/0550-3213(91)90064-5}{\texttt{
  doi:10.1016/0550-3213(91)90064-5}}.

\bibitem{vanNeerven:1991gh}
\hrefCMSnoop {} {W.~L. van Neerven and E.~B. Zijlstra, ``{The O($\alpha_S^{2}$)
  corrected Drell--Yan $K$-factor in the DIS and $\overline{\mathrm MS}$
  schemes}'',} \textit{ Nucl. Phys.} \textbf{ B382} (1992) 011. Erratum-ibid.
  {\bf B680} (2004) 513.
\href{http://dx.doi.org/10.1016/0550-3213(92)90078-P}{\texttt{
  doi:10.1016/0550-3213(92)90078-P}}.

\bibitem{ZWPROD}
\hrefCMSnoop {} {R.~Hamberg, T.~Matsuura, and W.~van Neerven}. {Z}WPROD program
  (1989-2002).

\bibitem{Mathews:2005bw}
\hrefCMSnoop {} {P.~Mathews, V.~Ravindran, and K.~Sridhar, ``{NLO-QCD}
  corrections to dilepton production in the {Randall-Sundrum} model'',}
  \textit{ JHEP} \textbf{ 10} (2005) 031,
  \href{http://www.arXiv.org/abs/hep-ph/0506158}{\texttt{
  arXiv:hep-ph/0506158}}.
\href{http://dx.doi.org/10.1088/1126-6708/2005/10/031}{\texttt{
  doi:10.1088/1126-6708/2005/10/031}}.

\bibitem{PTDR2}
\hrefCMSnoop {} {{CMS Collaboration}, ``{CMS technical design report, volume
  II: Physics performance}'',} \textit{ J. Phys.} \textbf{ G34} (2007) 995.
\href{http://dx.doi.org/10.1088/0954-3899/34/6/S01}{\texttt{
  doi:10.1088/0954-3899/34/6/S01}}.

\end{thebibliography}\endgroup

\cleardoublepage\appendix\section{The CMS Collaboration \label{app:collab}}\begin{sloppypar}\hyphenpenalty=5000\widowpenalty=500\clubpenalty=5000\textbf{Yerevan Physics Institute,  Yerevan,  Armenia}\\*[0pt]
S.~Chatrchyan, V.~Khachatryan, A.M.~Sirunyan, A.~Tumasyan
\vskip\cmsinstskip
\textbf{Institut f\"{u}r Hochenergiephysik der OeAW,  Wien,  Austria}\\*[0pt]
W.~Adam, T.~Bergauer, M.~Dragicevic, J.~Er\"{o}, C.~Fabjan, M.~Friedl, R.~Fr\"{u}hwirth, V.M.~Ghete, J.~Hammer\cmsAuthorMark{1}, S.~H\"{a}nsel, M.~Hoch, N.~H\"{o}rmann, J.~Hrubec, M.~Jeitler, G.~Kasieczka, W.~Kiesenhofer, M.~Krammer, D.~Liko, I.~Mikulec, M.~Pernicka, H.~Rohringer, R.~Sch\"{o}fbeck, J.~Strauss, F.~Teischinger, P.~Wagner, W.~Waltenberger, G.~Walzel, E.~Widl, C.-E.~Wulz
% ...
\vskip\cmsinstskip
% ...
1:~~Also at CERN, European Organization for Nuclear Research, Geneva, Switzerland\\
% ...

\end{sloppypar}
\end{document}